# Read before you cite!

M.V. Simkin and V.P. Roychowdhury
*Department of Electrical Engineering, University of California, Los Angeles, CA 90095-1594*

**Abstract.** We report a method of estimating what percentage of people who cited a paper had actually read it. The method is based on a stochastic modeling of the citation process that explains empirical studies of misprint distributions in citations (which we show follows a Zipf law). *Our estimate is only about 20% of citers read the original[1]*.

Many psychological tests have the so-called *lie-scale*. A small but sufficient number of questions that admit only one true answer, such as "Do you **always** reply to letters immediately after reading them?" are inserted among others that are central to the particular test. A wrong reply for such a question adds a point on the lie-scale, and when the *lie-score* is high, the over-all test results are discarded as unreliable. Perhaps, for a scientist the best candidate for such a lie-scale is the question "Do you read **all** of the papers that you cite?"

Comparative studies of the popularity of scientific papers has been a subject of much recent interest [1]-[8], but the scope has been limited to citation distribution analysis. We have discovered a method of estimating what percentage of people who cited the paper had actually **read** it. Remarkably, this can be achieved without any testing of the scientists, but solely on the basis of the information available in the ISI citation database.

Freud [9] had discovered that the application of his technique of Psychoanalysis to slips in speech and writing could reveal a lot of hidden information about human psychology. Similarly, we find that the application of statistical analysis to **misprints in scientific citations** can give an insight into the process of scientific writing. As in the Freudian case, the truth revealed is embarrassing. For example, an interesting statistic revealed in our study is that a lot of misprints are identical. Consider, for example, a 4-digit page number with one digit misprinted. There can be $10^4$ such misprints. The probability of repeating someone else's misprint accidentally is $10^{-4}$. There should be almost no repeat misprints by coincidence. One concludes that repeat misprints are due to copying some one else's reference, without reading the paper in question.

In principle, one can argue that an author might copy a citation from an unreliable reference list, but still read the paper. A modest reflection would convince one that this is relatively rare, and cannot apply to the majority. Surely, in the pre-internet era it took almost equal effort to copy a reference as to type in one's own based on the original, thus providing little incentive to copy if someone has indeed read, or at the very least has procured access to the original. Moreover, if someone accesses the original by tracing it from the reference list of a paper with a misprint, then with a high likelihood, the misprint has been identified and will not be propagated. In the past decade with the advent of the Internet, the ease with which would-be non-readers can copy from unreliable sources, as well as would-be readers can access the original has become equally convenient, but there is no increased incentive for those who read the original to also make verbatim copies, especially from unreliable resources[2]. In the rest of this paper, giving the benefit of doubt to potential non-readers, we adopt a much more generous view of a "reader" of a cited paper, as someone who at the very least consulted a trusted source (e.g., the original paper or heavily-used and authenticated databases) in putting together the citation list.

---

[1] Acknowledging the subjectivity inherent in what reading might mean to different individuals, we generously consider a "reader" as someone who has at the very least consulted a trusted source (e.g., the original paper or heavily-used authenticated databases) in putting together the citation list.

[2] According to many researchers the Internet may end up even aggravating the copying problem: more users are copying second-hand material without verifying or referring to the original sources.

As misprints in citations are not too frequent, only celebrated papers provide enough statistics to work with. Figure 1 shows distribution of misprints to one of such papers [10] in the rank-frequency representation, introduced by Zipf [11]. The most popular misprint in a page number propagated 78 times. Figure 2 shows the same data, but in the number-frequency format.

As a preliminary attempt, one can estimate an *upper bound* on the ratio of the number of readers to the number of citers, $R$, as the ratio of the number of **distinct** misprints, $D$, to the **total number** of misprints, $T$:[3]

$$R \approx D/T . \qquad (1)$$

Substituting $D = 45$ and $T = 196$ in Eq.(1), we obtain that $R \approx 0.23$. This estimate would be correct if the people who introduced original misprints had always read the original paper. However, given the low value of the upper bound on $R$, it is obvious that many original misprints were introduced while copying references. Therefore, a more careful analysis is needed. We need a model to accomplish it.

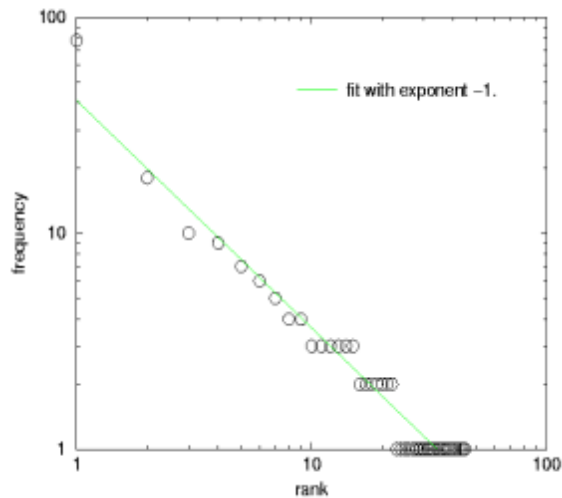

**Figure 1**. Rank-frequency distribution of misprints referencing a paper, which had acquired 4300 citations. There are 196 misprints total, out of which 45 are distinct. The most popular misprint propagated 78 times. A good fit to Zipf law is evident.

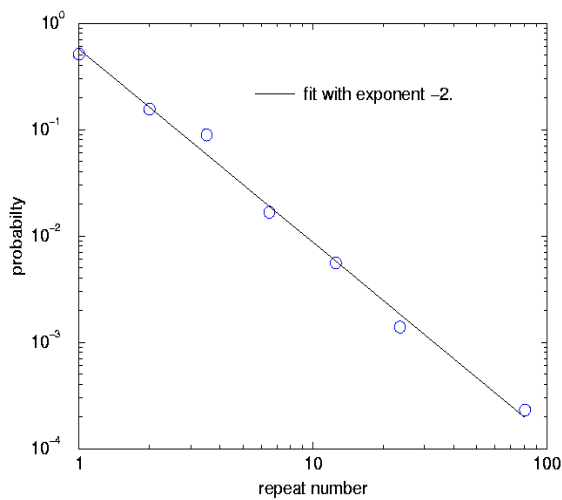

**Figure 2.** Same data as in Figure 1, but in the number-frequency representation. Misprints follow a power-law distribution with exponent close to 2.

---

[3] We know for sure that among $T$ citers, $T - D$ copied, because they repeated someone else's misprint. For the $D$ others, with the information at hand, we don't have any evidence that they did not read, so according to the presumed innocent principle, we assume that they read. Then in our sample, we have $D$ readers and $T$ citers, which leads to Eq.(1).

Our model for misprints propagation, which was stimulated by Simon's [12] explanation of Zipf Law and Krapivsky-Redner [4] idea of link redirection, is as follows. Each new citer finds the reference to the original in any of the papers that already cite it. With probability $R$ he reads the original. With probability $1-R$ he copies the citation to the original from the paper he found the citation in. In any case, with probability $M$ he introduces a new misprint.

The evolution of the misprint distribution (here $N_K$ denotes the number of misprints that propagated $K$ times, and $N$ is the total number of citations) is described by the following rate equations:

$$\frac{dN_1}{dN} = M - (1-R) \times (1-M) \times \frac{N_1}{N},$$

$$\frac{dN_K}{dN} = (1-R) \times (1-M) \times \frac{(K-1) \times N_{K-1} - K \times N_K}{N} \quad (K>1) \quad (2)$$

These equations can be easily solved using methods developed in [4] to get:

$$N_K \sim 1/K^g; \quad g = 1 + \frac{1}{(1-R) \times (1-M)}. \quad (3)$$

As the exponent of the number-frequency distribution, $g$, is related to the exponent of the rank-frequency distribution, $a$, by a relation $g = 1 + \frac{1}{a}$, Eq. (3) implies that:

$$a = (1-R) \times (1-M). \quad (4)$$

The rate equation for the total number of misprints is:

$$\frac{dT}{dN} = M + (1-R) \times (1-M) \times \frac{T}{N}, \quad (5)$$

The stationary solution of Eq. (5) is:

$$T = N \times \frac{M}{R + M - MR}. \quad (6)$$

The expectation value for the number of distinct misprints is obviously

$$D = N \times M. \quad (7)$$

From Equations (6) and (7) we obtain:

$$R = \frac{D}{T} \times \frac{N-T}{N-D}, \quad (8)$$

Substituting $D = 45$, $T = 196$, and $N = 4300$ in Equation (8), we obtain $R \approx 0.22$, which is very close to the initial estimate, obtained using Eq.(1). This low value of $R$ is consistent with the "Principle of Least Effort"[11].

One can ask why we did not choose to extract $R$ using Equations (3) or (4). This is because $a$ and $g$ are not very sensitive to $R$ when it is small. In contrast, $T$ scales as $1/R$.

We can slightly modify our model and assume that original misprints are only introduced when the reference is derived from the original paper, while those who copy references do not introduce new misprints (e.g. they do cut and paste). In this case one can show that $T = N \times M$ and $D = N \times M \times R$. As a consequence Eq.(1) becomes exact (in terms of expectation values, of course).

Preceding analysis assumes that the stationary state had been reached. Is this reasonable? Eq.(5) can be rewritten as:

$$\frac{d(T/N)}{M - (T/N) \times (R + M - M \times R)} = d \ln N. \quad (9)$$

As long as $M$ is small it is natural to assume that the first citation was correct. Then the initial condition is $N = 1; T = 0$. Eq.(9) can be solved to get:

$$T = N \times \frac{M}{R + M - M \times R} \times \left(1 - \frac{1}{N^{R+M-M \times R}}\right) \quad (10)$$

This should be solved numerically for $R$. For our guinea pig Eq. (10) gives $R \approx 0.17$.

Just as a cautionary note, Eq. (10) can be rewritten as:

$$\frac{T}{D} = \frac{1}{x} \times \left(1 - \frac{1}{N^x}\right); \quad x = R + M - M \times R \quad (11)$$

The definition of the natural logarithm is: $\ln a = \lim_{x \to 0} \frac{a^x - 1}{x}$. Comparing this with Eq. (11) we see that when $R$ is small ($M$ is obviously always small):

$$\frac{T}{D} \approx \ln N. \quad (12)$$

This means that a naïve analysis using Eq.(1) or Eq.(8) can lead to an erroneous belief that more cited papers are less read.

We conclude that misprints in scientific citations should not be discarded as mere happenstance, but, similar to Freudian slips, – **analyzed**.

We are grateful to J.M. Kosterlitz, A.V. Melechko, and N. Sarshar for correspondence.